\documentclass[twocolumn]{aastex631}
\usepackage{amsmath}
\usepackage{cases}
\usepackage{diagbox} 
\usepackage{makecell}
\begin{document}

\title{Effect of flow-aligned external magnetic fields on mushroom instability}

\author[0009-0009-7523-5887]{Yao Guo}
\affiliation{State Key Laboratory of Dark Matter Physics, Key Laboratory for Laser Plasmas, Department of Physics and Astronomy, Shanghai Jiao Tong University, Shanghai 200240, People’s Republic of China}

\author[0000-0001-5738-5739]{Dong Wu}
\affiliation{State Key Laboratory of Dark Matter Physics, Key Laboratory for Laser Plasmas, Department of Physics and Astronomy, Shanghai Jiao Tong University, Shanghai 200240, People’s Republic of China}
\affiliation{
Collaborative Innovation Center of IFSA (CICIFSA), Shanghai Jiao Tong University, Shanghai 200240, People’s Republic of China}

\author[0000-0001-7821-4808]{Jie Zhang}

\affiliation{State Key Laboratory of Dark Matter Physics, Key Laboratory for Laser Plasmas, Department of Physics and Astronomy, Shanghai Jiao Tong University, Shanghai 200240, People’s Republic of China}

\affiliation{
Collaborative Innovation Center of IFSA (CICIFSA), Shanghai Jiao Tong University, Shanghai 200240, People’s Republic of China}

\affiliation{Tsung-Dao Lee Institute, Shanghai Jiao Tong University, Shanghai 201210, People’s Republic of China}%

\correspondingauthor{Dong Wu}
\email{dwu.phys@sjtu.edu.cn}

\correspondingauthor{Jie Zhang}
\email{jzhang1@sjtu.edu.cn}



\begin{abstract}
Mushroom instability (MI) is a shear instability considered responsible for generating and amplifying magnetic fields in relativistic jets. While astrophysical jets are usually magnetized, how MI acts in magnetized jets remains poorly understood. In this paper, we investigate the effect of a flow-aligned external magnetic field on MI, with both theoretical analyses and particle-in-cell (PIC) simulations. In the limit of a cold and collisionless plasma, we derive a generalized dispersion relation for linear growth rates of the magnetized MIs. Numerical solutions of the dispersion relation reveal that the external magnetic field always suppresses the growth of MI, though MIs are much more robust against the external magnetic field than electron-scale Kelvin-Helmholtz instabilities (ESKHIs). Analyses are also extended to instabilities with an arbitrary wavevector in the shear interface plane, where coupling effect is observed for sub-relativistic scenarios.  Two-dimensional PIC simulations of single-mode MIs reach a good agreement with our analytical predictions, and we observe formation of a quasi-steady saturation structure in magnetized runs. In simulations with finite temperatures, we observe the competition and cooperation between MIs and a diffusion-induced DC magnetic field.
\end{abstract}


\keywords{Relativistic jets(1390) --- Plasma jets(1263) --- Plasma physics(2089) --- Magnetic fields(994)}


\section{Introduction} \label{sec:intro}
Relativistic jets have attracted broad interest as an important constituent of the gamma ray bursts (GRBs; \citet{Piran2005}), active galactic nuclei (AGNs; \citet{Blandford2019}), and many other astrophysical activities \citep{Mirabel1999}. The jets are considered to be sites of intricate physical processes, where the kinetic energy of the jet flow is dissipated into the energy of electromagnetic fields \citep{Pudritz2012}, which are capable of accelerating particles to superrelativistic velocities in turn \citep{Rieger2004} and triggering non-thermal emissions \citep{bottcher2007modeling}. However, the mechanism underlying the energy transition and subsequent processes remains debatable up-to-date, where many possible mechanisms have been proposed \citep{Weibel1959,Medvedev1999,Silva2003,Kulsrud2008,Alves2018}.

A possibility is that the dissipation processes are (partly) mediated by shear instabilities in the neighborhood of the jet boundary, where a sharp transverse velocity gradient exists between the jet and the ambient plasma. In the collisionless regime, 3D particle-in-cell (PIC) simulations showed that the jet boundary is unstable to electromagnetic instabilities \citep{Alves2012large}, where small fluctuations of electromagnetic fields are dramatically magnified by dissipating the kinetic energy of the shear flow, until being saturated. Theoretical analyses and 2D simulations conducted in separate planes have revealed that the electromagnetic fields are actually magnified by electron-scale Kelvin-Helmholtz instabilities (ESKHIs; \citet{Gruzinov2008}), mushroom instabilities (MIs; \citet{Alves2015Transverse}) and their coupling if in 3D scenarios. Both kinds of instabilities are dominated by electron (and also positron, if present) dynamics and take place on a timescale of  $1/\omega_{pe}$. While ESKHIs operate in the plane of the shear flow (with a wavevector parallel to the flow direction), MIs develop in planes perpendicular to the flow velocity (with a wavevector perpendicular to the shear plane). While ESKHIs are characterized with vortex structures of electrons, MIs exhibit a mushroom-like structure \citep{Alves2015Transverse} in their nonlinear stage, resembling the morphology of classical Rayleigh-Taylor instabilities. Particularly, \citet{Alves2015Transverse} showed that while the maximum linear growth rate of MI ($\sigma_{\text{max}}=\omega_{pe}\gamma_0V_0/c$ for a symmetrical shear, where $V_0$ is the shear velocity and $\gamma_0$ is the corresponding Lorentz factor) is smaller than that of ESKHI ($\sigma_{\text{max}}=\omega_{pe}/2\sqrt{2}$) for subrelativistic shear velocities ($V_0< c/3$), instead it is much larger than the latter in the relativistic limit, indicating the dominance of MI in relativistic jets. Recently, \citet{Kawashima2022} proposed that MI can drive magnetic reconnection, which can trigger particle acceleration and lead to jet spine formation, and both are characteristic problems in jets. Therefore, investigating the behavior of MI is significant for us to understand the mechanism of jets.

Since the magnetization process of the jet depends on specific scenarios and remains not so clear, it is undetermined whether the jet has been (partly) magnetized or not when MI begins to operate efficiently. While the latter scenario has been investigated in detail both theoretically and with PIC simulations \citep{Alves2015Transverse,Yao2020}, the former scenario is much less studied and understood. Related research used PIC simulations to study the effect of an external magnetic field on jets  \citep{Dieckmann2019,Amin2024}\footnote{We note that \citet{Amin2024} discussed the effect of a toroidal external magnetic field, but their expression for the field seems to lack a minus sign to be actually toroidal.}, concluding the stabilizing effect of a flow-aligned external magnetic field on the jet (and the instability). However, an analytical interpretation of this result has remained missing.

On the other hand, the effect of an external magnetic field on ESKHIs has been well studied. In the incompressible (and hence non-relativistic) limit, \citet{Che2023,Che2025} studied the effect of an external magnetic field in an arbitrary direction and derived an explicit dispersion relation for magnetized ESKHI growth rates. In the cold but relativistic regime, \citet{Guo2025} derived a generalized eigenequation in the case of a flow-aligned external magnetic field, and also revealed interesting interplay among the external field, ESKHIs, and the DC (direct current) magnetic field, with PIC simulations. 

It is noteworthy that, however, MIs are theoretically stable in the incompressible limit (see \citet{Che2023}, where $\sigma\propto \textbf{\textit{k}}\cdot \textbf{\textit{V}}$ in their dispersion relation, while in MI $\textbf{\textit{k}}\cdot \textbf{\textit{V}}=0$), while the cold and relativistic equations give an explicit description of them \citep{Alves2015Transverse}. Given that the system of equations related to the 3D dynamics of the cold and magnetized shear flow are already derived in \citet{Guo2025}, it is straightforward to extend the analyses to contain MI and also its coupling with ESKHI.

In this paper, we study the effect of a flow-aligned external magnetic field on MI with both theoretical analyses and PIC simulations. In Sec. \ref{sec:theo}, we give a description of the problem and derive a generalized eigenequation for linear growth rates of MIs. Numerical solutions of the dispersion relation are also presented. In Sec.\ \ref{sec:sim}, we present and analyze the results of our PIC simulations to verify the theoretical results. Simulations are carried out either in cold plasmas with single-mode perturbations, or in thermal plasmas with no initial perturbation. In Sec.\ \ref{sec:conclusion}, we draw a conclusion and present some discussions. The mathematical details of deriving the eigenequation are presented in Appendix \ref{app:3D}. We use SI units throughout the paper.

\section{Theoretical analyses} \label{sec:theo}

In this section, we formulate the MI dispersion relation by following the analytical approach in \citet{Guo2025}, which was used to study the effect of a flow-aligned external magnetic field on ESKHI. The methodology is outlined as follows. 

\subsection{Physical settings of the magnetized shear flow}\label{sec:setting}

We consider a cold and collisionless plasma consisting of only electrons and protons. We utilize the relativistic fluid equations in the limit of cold and collisionless plasma:
\begin{align}
&\frac{\partial n}{\partial t}+\nabla\cdot(n\textbf{\textit{v}})=0, \label{eq:1}\\
&\left(\frac{\partial}{\partial t}+\textbf{\textit{v}}\cdot\nabla \right)\textbf{\textit{p}}+e\left(\textbf{\textit{E}}+\textbf{\textit{v}}\times \textbf{\textit{B}}\right)=0,    \label{eq:2}
\end{align}
where the momentum is $\textbf{\textit{p}}=\gamma m \textbf{\textit{v}}$ and the Lorentz factor is $\gamma=1/\sqrt{1-v^2/c^2}$. These equations are coupled with the Maxwell equations
\begin{align}
&\nabla\times \textbf{\textit{E}}=-\frac{\partial \textbf{\textit{B}}}{\partial t},\label{eq:3}\\
&\nabla\times \textbf{\textit{B}}=-\mu_0\textbf{\textit{J}}+\frac{1}{c^2}\frac{\partial \textbf{\textit{E}}}{\partial t},\label{eq:4}
\end{align}
with the electric current $\textbf{\textit{J}}$ being 
\begin{equation}
\textbf{\textit{J}}=-en_e\textbf{\textit{v}}_e+en_i\textbf{\textit{v}}_i. \label{eq:5}
\end{equation}
Now we consider a shear flow with a non-uniform velocity distribution $\textbf{\textit{v}}_{e0}=\textbf{\textit{v}}_{i}=[0,v_0(x),0]$. Still, the ions are considered to be free-streaming and unperturbed throughout the timescale we investigate. The initial number density distribution is $n_{e0}(x)=n_i(x)$, maintaining the charge and current neutrality of the shear flow.  A flow-aligned external magnetic field $\textbf{\textit{B}}_0=[0,B_0(x),0]$ is applied to the system. In the cold plasma with thermal pressure vanishing, the dynamical equilibrium of ions and electrons requires $ \textbf{\textit{E}}_0=0$, given $ \textbf{\textit{v}}_{e0}\times  \textbf{\textit{B}}_0=0$. No net current $\textbf{\textit{J}}_{0}$ results in a uniform magnetic field $B_0(x)=\text{const.}$ We note that uniform external magnetic fields with components in all three directions were considered by \citet{Che2025} for ESKHIs in incompressible plasmas; However, this situation will bring much complexity and especially discontinuities in our sets of equations, leading us to start with the simple case of a flow-aligned field.

\subsection{The eigenequation and the dispersion relation of MI linear growth rate}\label{sec:eigen}

MI is characterized with a wavevector $\textbf{\textit{k}}=k_z\textbf{\textit{e}}_z$ perpendicular to the shear flow, while $\textbf{\textit{k}}=k_y\textbf{\textit{e}}_y$ for ESKHIs. In order to analyze the dynamics of MIs, we consider perturbations in the form of $f(x,z,t)=\tilde{f}(x)e^{i(k_zz-\omega t)}$, while the whole system is assumed to be uniform in the $y$ direction. Linearize Eqs.\ \ref{eq:1}$\sim$\ref{eq:5} with the perturbed variables, reduce them until only two perturbed variables $\tilde E_y(x)$ and $\tilde E_z(x)$ remain, and then we can derive the two coupled eigenequations for linear growth rates of magnetized MI: [See Appendix \ref{app:3D} for the linearization process of the equations, where the perturbations are generalized to $f(x,y,z,t)=\tilde{f}(x)e^{i(k_yy+k_zz-\omega t)}$, encompassing MIs and ESKHIs simultaneously in the eigenequations.]

\begin{widetext}
\begin{eqnarray}
       &&\left[\frac{\omega^2(\omega_c^2+\omega_p^2-\omega^2)+k_z^2c^2(\omega^2-\omega_c^2+(\gamma_0^2-1)\omega_p^2)}{D(\omega,k_z)}\tilde E_y'\right]'
       -\left[\frac{k_z\omega_c\omega_p^2((1-\gamma_0^2)\omega^2+\gamma_0^2k_z^2v_0^2)}{\omega D(\omega,k_z)}\right]'\tilde E_y +\left[\frac{k_zv_0\omega \gamma_0^2\omega_p^2}{D(\omega,k_z)}\tilde E_z'\right]'  +\nonumber\\
         &&    \left[\frac{\omega_p^2-\omega^2}{c^2}+\frac{k_z^2v_0^2\gamma_0^2\omega_p^2}{c^2(\omega^2-\omega_c^2)}+k_z^2\right]\tilde E_y+ \left[\frac{v_0\omega_c\gamma_0^2\omega_p^2\left(\omega^2-k_z^2c^2\right)}{D(\omega,k_z)}\tilde E_z\right]'+
       \frac{k_z^2v_0\omega_c\gamma_0^2\omega_p^2}{D(\omega,k_z)}\tilde E_z'+
       \left[\frac{\omega k_zv_0\gamma_0^2\omega_p^2}{c^2(\omega^2-\omega_c^2)}\right] \tilde E_z=0 ,\label{eq:dis1} \\
   && \left[\left(1+\frac{c^2k_z^2(\omega^2-\omega_c^2)}{D(\omega,k_z)}\right)\tilde E_z'\right]'
    +\left[\frac{k_z\omega\omega_c\gamma_0^2\omega_p^2}{D(\omega,k_z)}\right]'\tilde E_z \nonumber+
    \frac{\omega^2[\omega^4+\gamma_0^4\omega_p^4+k_z^2c^2(\omega_c^2-\omega^2+\gamma_0^2\omega_p^2)-\omega^2(\omega_c^2+2\gamma_0^2\omega_p^2)]}{c^2D(\omega,k_z)}\tilde E_z -\nonumber\\
    &&\left[\frac{k_zv_0\omega\gamma_0^2\omega_p^2}{D(\omega,k_z)}\tilde E_y'\right]'
    +\left[\frac{k_z^2v_0\omega_c\gamma_0^2\omega_p^2}{D(\omega,k_z)}\tilde E_y\right]' +\frac{v_0\omega_c\gamma_0^2\omega_p^2\left(\omega^2-k_z^2c^2 \right)}{c^2D(\omega,k_z)}\tilde E_y'
    +\left[\frac{k_zv_0\omega\gamma_0^2\omega_p^2(c^2k_z^2-\omega^2+\gamma_0^2\omega_p^2)}{c^2D(\omega,k_z)}\right]\tilde E_y=0,  \label{eq:dis2}
\end{eqnarray}
where $\gamma_0=1/\sqrt{1-v_0^2/c^2}$ is the bulk Lorentz factor, $\omega_c=eB_0/\gamma_0 m_e$ is the electron cyclotron frequency under the external magnetic field, $\omega_p=\sqrt{n_{e0} e^2/m_e\epsilon_0\gamma_0^3}$ is the electron plasma frequency (we note that the normalization with respect to $\gamma_0$ can be different in other works), $D(\omega,k_z)=\omega^2\left(\omega^2-k_z^2c^2-\gamma_0^2\omega_p^2\right)-\omega_c^2(\omega^2-k_z^2c^2)$, and they are all function of $x$ due to dependence on $v_0(x)$ and $n_{e0}(x)$. The prime notation ``$'$" represents the partial derivative with respect to the $x$ coordinate. Equations\ \ref{eq:dis1} and \ref{eq:dis2} are generic and not confined to specific spatial profiles of the zeroth-order variables, except for a uniform $B_0$. In principle, with Eqs.\ \ref{eq:dis1} and \ref{eq:dis2} one can investigate the stability of a system with arbitrary profiles of $v_0(x)$ and $n_{e0}(x)$. Any imaginary (or complex) solution of the eigenfrequency $\omega$ represents an unstable mode of the system, with $\sigma=\text{Im}(\omega)$ being the linear instability growth rate.

If the instability is driven by a discontinuous shear flow [e.g. $v_0(x)=\text{sgn}(x)V_0$], we can integrate Eqs.\ \ref{eq:dis1} and \ref{eq:dis2} across the shear interface to derive the connecting boundary conditions:
\begin{eqnarray}
&& \left[\frac{\omega^2(\omega_c^2+\omega_p^2-\omega^2)+k_z^2c^2(\omega^2-\omega_c^2+(\gamma_0^2-1)\omega_p^2)}{D(\omega,k_z)}\tilde E_y'\right]\bigg|^{0_+}_{0_-}- \left[\frac{k_z\omega_c\omega_p^2((1-\gamma_0^2) \omega^2+\gamma_0^2k_z^2v_0^2)}{\omega D(\omega,k_z)}\tilde E_y\right]\bigg|^{0_+}_{0_-}+\nonumber\\
&&\left[\frac{k_zv_0\omega \gamma_0^2\omega_p^2}{D(\omega,k_z)}\tilde E_z'\right]\bigg|^{0_+}_{0_-}
+\left[\frac{v_0\omega_c\gamma_0^2\omega_p^2\left(\omega^2-k_z^2c^2\right)}{D(\omega,k_z)}\tilde E_z\right]\bigg|^{0_+}_{0_-}=0,\label{eq:dis10} \\ 
&&    \left[\left(1+\frac{c^2k_z^2(\omega^2-\omega_c^2)}{D(\omega,k_z)}\right)\tilde E_z'\right]\bigg|^{0_+}_{0_-}
    +\left[\frac{k_z\omega\omega_c\gamma_0^2\omega_p^2}{D(\omega,k_z)}\tilde E_z\right]\bigg|^{0_+}_{0_-}
    +\left[\frac{k_zv_0\omega\gamma_0^2\omega_p^2}{D(\omega,k_z)}\tilde E_y'\right]\bigg|^{0_+}_{0_-} +\left[\frac{k_z^2v_0\omega_c\gamma_0^2\omega_p^2}{D(\omega,k_z)}\tilde E_y\right]\bigg|^{0_+}_{0_-} =0 .  \label{eq:dis20}
\end{eqnarray}
\end{widetext}
In the unmagnetized scenario where  $\omega_c=0$, Eqs.\ \ref{eq:dis1}$\sim$\ref{eq:dis20} reduce to the set of equations derived by \citet{Alves2015Transverse}, which can be solved analytically to give the MI growth rate $\sigma$ for an unmagnetized symmetrical shear flow with $v_0(x)=\text{sgn}(x)V_0$ and $n_{e0}(x)=const.$:
\begin{equation}
    \frac{\sigma}{\omega_p}=\frac{1}{\sqrt{2}}\left[\sqrt{\frac{4k_z^2V_0^2\gamma_0^2}{\omega_p^2}+G(k_z)^2}-G(k_z)\right]^{1/2},
    \label{eq:alves}
\end{equation}
where $G(k_z)=(1+k_z^2c^2/\omega_p^2)$. This equation predicts the maximum growth rate at an infinite wavenumber: $\sigma_\text{max}=\sigma(k_z\xrightarrow[]{}\infty)=\gamma_0V_0\omega_p/c$. In contrast, when $\omega_c\neq0$, the equations are highly coupled and have to be solved numerically. When the shear flow is uniform in each half of the plane, one can assume solutions with the form of $\tilde E_y^{\pm}=A_1^{\pm}\exp(a^{\pm}x)+A_2^{\pm}\exp(b^{\pm}x)$ and $\tilde E_z^{\pm}=B_1^{\pm}\exp(a^{\pm}x)+B_2^{\pm}\exp(b^{\pm}x)$ [$F(x)=F^+$ at $x>0$ and $F(x)=F^-$ at $x<0$ respectively], where the real part of $a^+$ and $b^+$ are restricted to be negative, and that of $a^-$ and $b^-$ positive, guaranteeing a mode vanishing at infinities. Thus, Eqs.\ \ref{eq:dis1} and \ref{eq:dis2} are transformed into polynomial equations of $A_i^\pm,B_i^\pm,a^\pm$,$b^\pm$, and also $\omega$. Along with the boundary conditions of the electric field, the equations can be solved straightforward to give $\omega$ (see the Appendix in \citet{Guo2025} for detail).

\begin{figure}[h]
\centering
\includegraphics[width=0.98 \columnwidth]{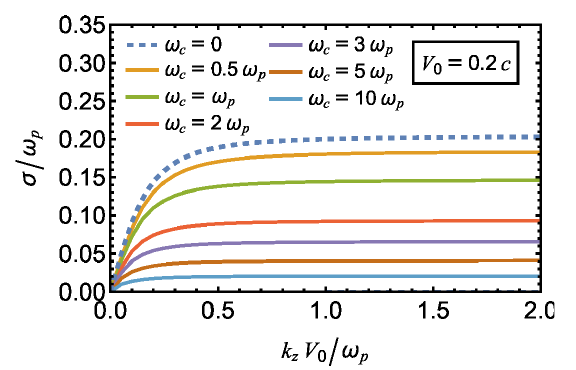}
\caption{The dispersion relation of the MI growth rate $\sigma(k_z)$ with $v_0=V_0 \text{sgn}(x)$ and $V_0=0.2c$ under uniform external magnetic field with different magnitudes $B_0$ corresponding to $\omega_c/\omega_p=eB_0/m_e\gamma_0\omega_p=$0.5 (yellow), 1 (green), 2 (red), 3 (purple), 5(brown), and 10 (cyan). Each curve is interpolated by numerically solving the eigenequation at intervals of $\Delta k_zV_0=\omega_p/20$. The theoretical dispersion relation in absence of external magnetic field, Eq.\ \ref{eq:alves} (blue and dashed), is also shown for comparison.}
\label{fig:dis02v}
\end{figure}

\begin{figure}[h]
\centering
\includegraphics[width=0.98 \columnwidth]{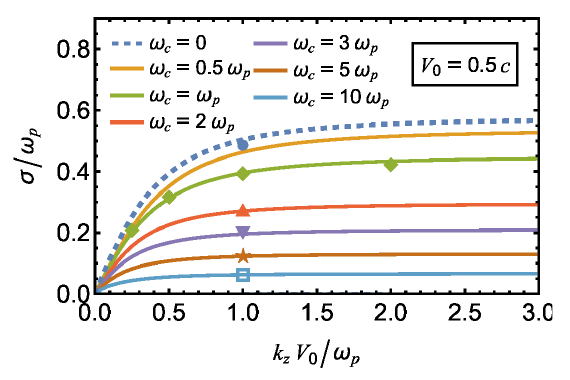}
\caption{The same plot as Fig.\ \ref{fig:dis02v}, but with $V_0=0.5c$. Results extracted from single-mode PIC simulations are also marked for comparison, where the horizontal coordinate of the marks corresponds to the wavenumber of the pre-imposed perturbation, and the vertical coordinate corresponds to the growth rates extracted from the simulations. The details of the simulations will be discussed in Sec.\ \ref{sec:sim}.}
\label{fig:dis05v}
\end{figure}

\begin{figure}[h]
\centering
\includegraphics[width=0.98 \columnwidth]{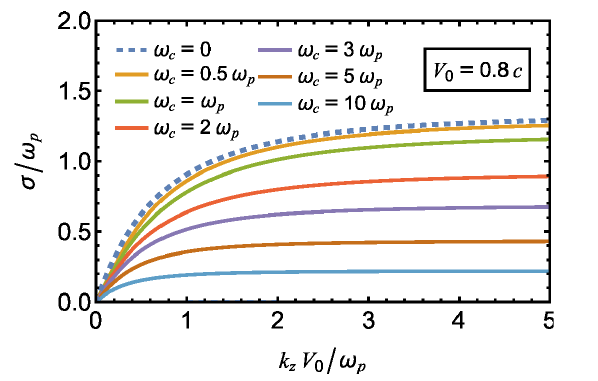}
\caption{The same plot as Fig.\ \ref{fig:dis02v}, but with $V_0=0.8c$.}
\label{fig:dis08v}
\end{figure}

We numerically solve MI growth rates $\sigma$ for symmetrical shear flows [$n_{e0}(x)=const. , v_0(x)=\text{sgn}(x)V_0,$ and $\omega_c(x)=eB_0/\gamma_0m_e$] with certain values of $V_0$ and $\omega_c/\omega_p$. Figs.\ \ref{fig:dis02v}, \ref{fig:dis05v}, and \ref{fig:dis08v} show the dispersion relation $\sigma(k_z)$ for $V_0/c=0.2$, 0.5 and 0.8 respectively. Especially, growth rates extracted from single-mode PIC simulations are also marked in Fig.\ \ref{fig:dis05v}, which will be discussed in detail in Sec.\ \ref{sec:sim}.
From Figs.\ \ref{fig:dis02v}$\sim$ \ref{fig:dis08v} we can conclude that: (1) Quite generally, the flow-aligned external magnetic fields suppress MI. The larger the $B_0$ (and hence $\omega_c$), the smaller the growth rate $\sigma$. We note that this is not the case for ESKHIs, which can be destabilized by the external magnetic field with relatively large values of $V_0$ or $\gamma_0$ \citep{Guo2025}.
(2) With same values of $\omega_c/\omega_p$, MIs driven by a larger shear velocity $V_0$ are more robust against the external magnetic field. 
(3) With larger magnitudes of $B_0$, the growth rate $\sigma(k_z)$ approaches the maximum value $\sigma(k_z\xrightarrow{}\infty)$ faster in $k_z$. Since thermal effects can bring in a cutoff at larger $k_z$ in the dispersion relation \citep{Alves2015Transverse}, this might indicate a smaller wavenumber $k_z$ where the most unstable mode with $\sigma_{\text{max}}$ occurs in realistic scenarios.

\subsection{Competition between MI and ESKHI}
Since MIs and ESKHIs exist simultaneously in shear flows, it is meaningful to investigate how they compete in shear flows with different conditions. In the unmagnetized scenario, the ESKHI growth rate of a symmetrical shear flow is \citep{Gruzinov2008}
\begin{equation}
    \frac{\sigma}{\omega_p}=\frac{1}{\sqrt 2}\left(\sqrt{1+8\frac{k_y^2V_0^2}{\omega_p^2}}-1-2\frac{k_y^2V_0^2}{\omega_p^2}\right)^{1/2}, \label{eq:gru}
\end{equation}
giving a maximum growth rate $\sigma_\text{max}=\omega_p/2\sqrt2$. Since the MI maximum growth rate is $\sigma_\text{max}=\gamma_0V_0\omega_p/c$, this indicates the dominance of ESKHIs in shear flows with $V_0<c/3$ and the dominance of MIs with $V_0>c/3$.

\begin{figure}[h]
\centering
\includegraphics[width=0.98 \columnwidth]{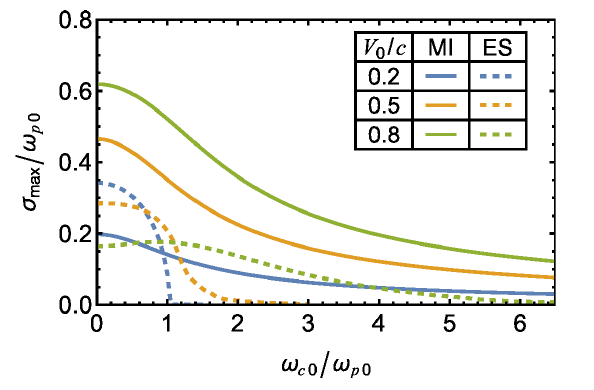}
 \caption{Normalized maximum growth rates $\sigma_\text{max}/\omega_{p0}$ of MIs (solid lines) and ESKHIs (dashed lines) as functions of the magnitude of the normalized external magnetic field $\omega_{c0}/\omega_{p0}=eB_0/m_e\omega_{p0}$, with $V_0/c=$0.2 (blue), 0.5 (yellow) and 0.8 (green). For consistent  comparison, here $\sigma_\text{max}$ and $B_0$ are normalized with the non-relativistic frequencies $\omega_{c0}=eB_0/m_e$ and $\omega_{p0}=\sqrt{n_{e0} e^2/m_e\epsilon_0}$, which only depend on $B_0$ and $n_{e0}$ respectively and remain invariant with different $V_0$ or $\gamma_0$. They are related to our previously defined relativistic frequencies by $\omega_c=\omega_{c0}/\gamma_0$ and $\omega_p=\omega_{p0}/\gamma_0^{3/2}$.}
\label{fig:com}
\end{figure}

The magnetized scenario can be investigated readily with our eigenequations. Figure\ \ref{fig:com} compares the maximum growth rates $\sigma_\text{max}$ of MIs and ESKHIs under different magnitudes of the external magnetic field $B_0$ (and hence $\omega_c$) and different shear velocities $V_0$. The MI growth rates are computed by solving Eqs.\ \ref{eq:dis1}$\sim$\ref{eq:dis20} with a large enough $k_z$, while the ESKHI growth rates are computed by setting $k_z=0$ and varying $k_y$ for a maximum solution of $\sigma$ in Eqs.\ \ref{eq:disa1}$\sim$\ref{eq:conna2}. We note that in Fig.\ \ref{fig:com} $\sigma_\text{max}$ and $B_0$ are normalized with the non-relativistic frequencies $\omega_{c0}=eB_0/m_e$ and $\omega_{p0}=\sqrt{n_{e0} e^2/m_e\epsilon_0}$, which are invariant with different $\gamma_0$, enabling direct comparison between cases with different shear velocities $V_0$. 

For all values of $V_0$ considered, it is obvious that MIs are more robust against the external magnetic field than ESKHIs. For each value of $V_0$, ESKHIs are completely stabilized by external magnetic fields larger than a threshold value. In contrast, MIs remain unstable under large values of $\omega_{c0}/\omega_{p0}$. In consequence, shear flows with $V_0/c<1/3$ should be dominated by ESKHIs under weak external magnetic fields, but dominated by MIs under strong external magnetic fields (see the crossing of the solid and dashed blues curves for $V_0=0.2c$ in Fig.\ \ref{fig:com}). Our results also imply that a finite threshold stabilizing magnetic field may not exist for MIs in the cold plasma limit. 

Although in the case with a larger $V_0/c$, the external magnetic field can be destabilizing for ESKHIs (e.g. $V_0/c=0.8$, $\omega_{c0}/\omega_{p0}\lesssim1.1$, see the green dashed curve in Fig.\ \ref{fig:com}), yet ESKHIs are not strong enough to compete with MIs, due to their different dependence on $\gamma_0$. In consequence, jets with relativistic velocities should be dominated by MIs whatever the magnetization.

\subsection{The dispersion relation of the coupled instability}

The discussion in the preceding subsection is an oversimplification where coupling between MIs and ESKHIs is neglected. In realistic scenarios, MIs and ESKHIs can actually be coupled in 3D dynamics of particle motion and electromagnetic fields. This coupled instability can be investigated directly by assuming perturbations in the form of $f(x,y,z,t)=\tilde{f}(x)e^{i(k_yy+k_zz-\omega t)}$, as detailed in Appendix \ref{app:3D}. By numerically solving Eqs.\ \ref{eq:disa1}$\sim$\ref{eq:conna2}, we can derive the dispersion function of the coupled instability linear growth rate $\sigma(k_y,k_z)$.

\begin{figure}[h]
\centering
\includegraphics[width=0.98 \columnwidth]{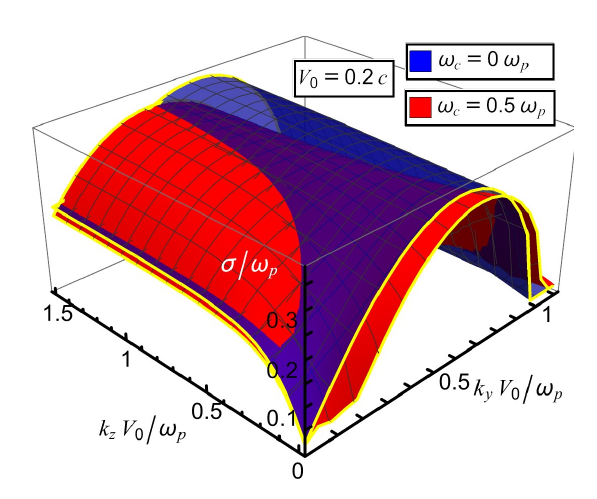}
 \caption{The dispersion relation of the coupled shear instability linear growth rate $\sigma(k_y,k_z)$, in a symmetrical shear flow with $V_0=0.2c$, under two different magnitudes of external magnetic field $\omega_c/\omega_p=$ 0 (blue) and 0.5 (red).}
\label{fig:3D}
\end{figure}

Figure\ \ref{fig:3D} shows the numerical solution of $\sigma(k_y,k_z)$ in a symmetrical shear flow with $V_0=0.2c$. The dispersion relation of the coupled instability inherits some characteristics from MI and ESKHI. Like MI, the unstable region of the coupled instability extends to an infinite $k_z$; On the other hand, the unstable region is cut off at some $k_y$, which is the same as the dispersion relation of ESKHI. In the unmagnetized case with $\omega_c=0$ (the blue surface), the dispersion relation largely resembles that of ESKHI dispersion relation (the yellow curve at $k_z=0$) extended over $k_z$, with the maximum growth rate slightly increasing with $k_z$. Interestingly, under the external magnetic field $\omega_c=0.5\omega_p$ (the red surface), the dispersion relation is distorted towards the $k_y=0$ axis as $k_z$ increases. Especially, the region with $0.05\lesssim k_yV_0/\omega_p\lesssim0.4$ and $k_zV_0/\omega_p\gtrsim0.4$ is destabilized in this case (where the red surface is above the blue one), while both ESKHIs and MIs in separate are stabilized by this magnetic field.

We also investigated the dispersion relation of the coupled instability with larger $V_0/c=$ 0.5 and 0.8. With larger $V_0$, the coupling is less important, since the instability is largely dominated by MI, with the maximum growth rates achieved at $k_y=0$ and $k_z\xrightarrow[]{}\infty$. Also, the instability is generally suppressed by the external magnetic field, as is MI, unless at large $k_y$. Although the unstable region is extended to larger $k_y$ under the external field, growth rates there are much smaller compared to the maximum growth rate at $k_y=0$. In consequence, in relativistic and cold ($V_0\gtrsim0.5c\gg v_{th}$) jets, the instability at the shear interface should resemble the quasi-2D morphology of a pure MI.

\section{PIC simulations} \label{sec:sim}
Simulating MI requires a self-consistent description of the electron dynamics and electromagnetic fields at least. In order to verify our analytical results and maintain kinetic effects, we carry out 2D3V simulations with a fully kinetic PIC code LAPINS \citep{Wu2019}. In consistency with our theoretical framework, the simulations are carried out in the $xOz$ plane, while velocities of particles and electromagnetic fields are allowed to have components in all the directions. Periodic boundary conditions are applied to the simulation domain in all directions. The initial number density distribution is uniform, with $n_{e0}=n_i=10^{15}\ m^{-3}$. The simulation domain is $L_x\times L_z$, and both particle species share an initial sheared velocity profile:
    \begin{equation}
    v_{y}(x)=\begin{cases}
    -V_0,& \text{if } x < L_x/4, \\
    V_0,& \text{if } L_x/4\leq x \leq 3L_x/4,\\
    -V_0,& \text{if } x > 3L_x/4.
    \end{cases}
    \label{eq:velocity}
\end{equation}
An external magnetic field $B_0$ is also applied in the $y$ direction. We here investigate the case with $V_0=0.5c$, which results in a $\gamma_0\approx1.15$ and $\omega_p\approx1.44\times10^9\ \text{s}^{-1}$.

\subsection{Single-mode simulations}

According to Eq.\ \ref{eq:alves}, in the cold plasma limit, the maximum growth rate of MI is achieved at $k_z\rightarrow {}\infty$ and $\lambda_z \rightarrow{}0$, which is unrealistic in simulations due to the finite length of grid cells. In consequence, here we first investigate the evolution of single-mode instabilities with certain wavelengths, to give a direct verification of the analytical dispersion relation. 

In simulations, we trigger a single-mode instability by imposing a very small initial velocity perturbation $v_{x0}=0.0002c \cos(2\pi z/L_z)$ on the electrons in a cold plasma, where the initial thermal velocity of both species is $v_{the}=v_{thi}=0$. We conduct two groups of simulations, with the first fixing the magnitude of the external magnetic field with $\omega_c=eB_0/m_e\gamma_0=\omega_p$, and varying the wavenumber $k_z$ (hence the simulation domain width $L_z=2\pi/k_z$) with $k_zV_0/\omega_p=2\pi V_0/L_z\omega_p= 0.25,0.5,1$ and 2; the second group fixing $k_zV_0/\omega_p=1$, and varying $\omega_c/\omega_p=0,1,2,3,5$ and 10.  $L_x$ is restricted to be $4\pi c/\omega_p$, and the simulation domain is divided by square grid cells with $\delta_x=\delta_z=L_x/800=\pi c/200\omega_p$. Each grid cell is filled with 100 macro-particles per species.

\begin{figure}[h]
\centering
\includegraphics[width=0.98 \columnwidth]{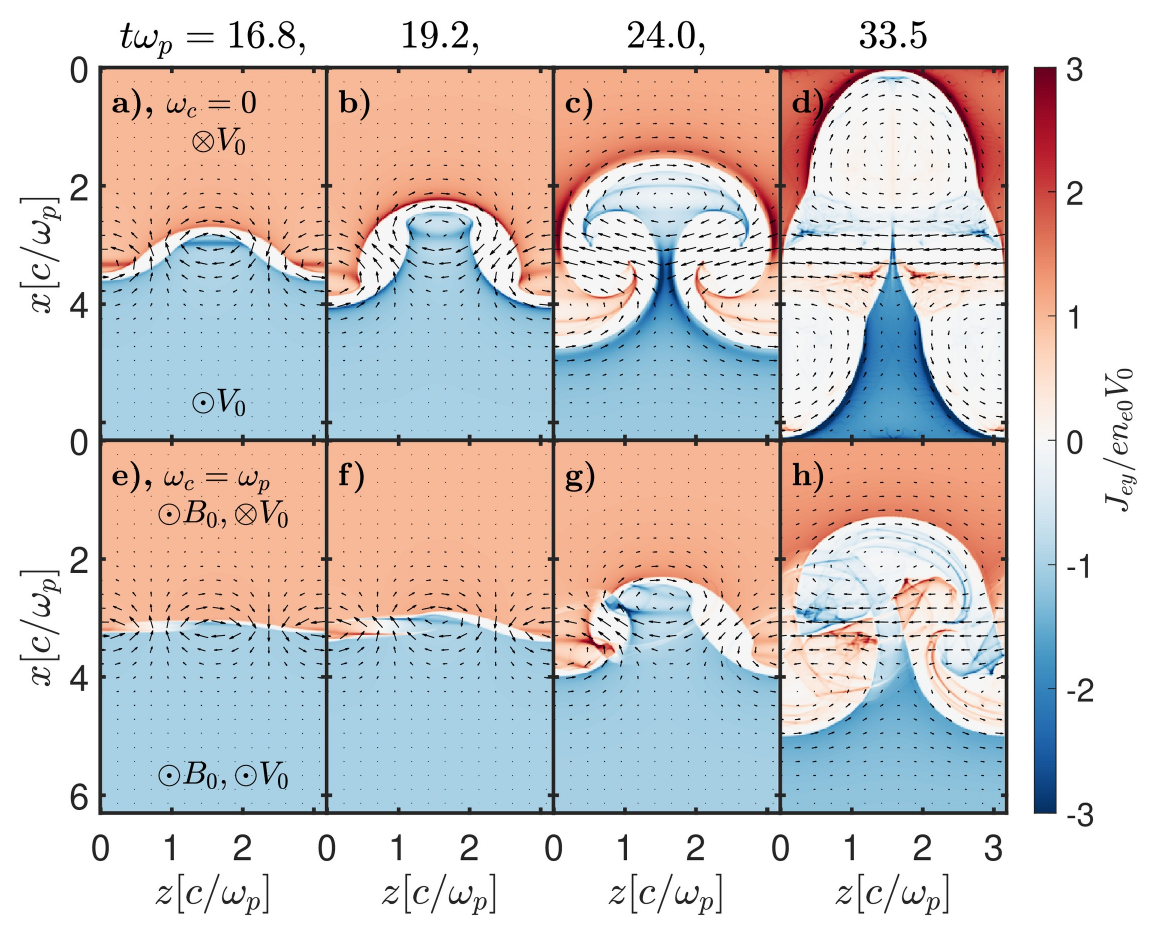}
 \caption{The evolution of the electron current $J_{ey}$, in two different $kV_0=\omega_p$ runs with $\omega_c=0$ [Panels (a)$\sim$(d)] and $\omega_c=\omega_p$ [Panels (e)$\sim$(h)], at four simulation times $t\omega_p=$ 16.8 (the first column), 19.2 (the second column), 24.0 (the third column) and 33.5 (the fourth column). The direction of the external magnetic field and the initial velocity in the runs is marked in Panels (a) and (e). The in-plane magnetic field generated by MI is displayed as black arrows.  Only half of the simulation domain $0<x<L_x/2$ is shown.}
\label{fig:Jx}
\end{figure}

Figure \ref{fig:Jx} shows the temporal evolution of $J_{ey}$ in two $kV_0=\omega_p$ runs, with $\omega_c/\omega_p=0$ (the first row) and 1 (the second row) respectively. Generally, the behavior of MI can be understood as follows\ \citep{Kawashima2022}. Panels (a)/(e)$\sim$(b)/(f): (1) Since the perturbed electrons cross the shear interface and the protons remain still, a sinusoidal current sheet of $J_y$ develops along the shear interface. (2) The sinusoidal current sheet generates a magnetic field with a dominant $B_z$ component on either side of the perturbation crests (while $B_x$ dominates between the crests). (3) The $B_z$ introduces a Lorentz force in the $x$ direction on the bulk of electrons, which has a streaming velocity $v_y=\pm V_0$. (4) The Lorentz force displaces the electrons on one side across the shear interface, and push electrons on the other side away from the interface. This amplifies the deformation of the interface, thereby reinforcing the instability feedback loop. (5) As the magnetic field grows, the electrons are pushed away from the shear interface by the magnetic pressure, forming a density gap along the interface. Panel (c)/(g): (6) When the instability enters the nonlinear stage, the electric field at the edge of the electron bulk induces it to roll up, forming the ``mushroom'' structure. Panel (d)/(h): (7) As the deformation of the interface becomes greater, a strong DC component of $B_z$ forms along the shear interface, cutting the ``mushroom cup'' off from the electron bulk and the instability finally saturates.

Comparing the two runs in Figure \ref{fig:Jx}, it is evident that MI grows slower in magnetized shear flows, in agreement with our analytical prediction. In the magnetized run, as the electrons cross the interface with a velocity $v_x$, the ``mushroom'' is distorted by the Lorentz force $F_z=-v_xB_0$, leading to asymmetry on both sides, see Fig.\ \ref{fig:Jx}(g). 
\begin{figure}[h]
\centering
\includegraphics[width=0.98 \columnwidth]{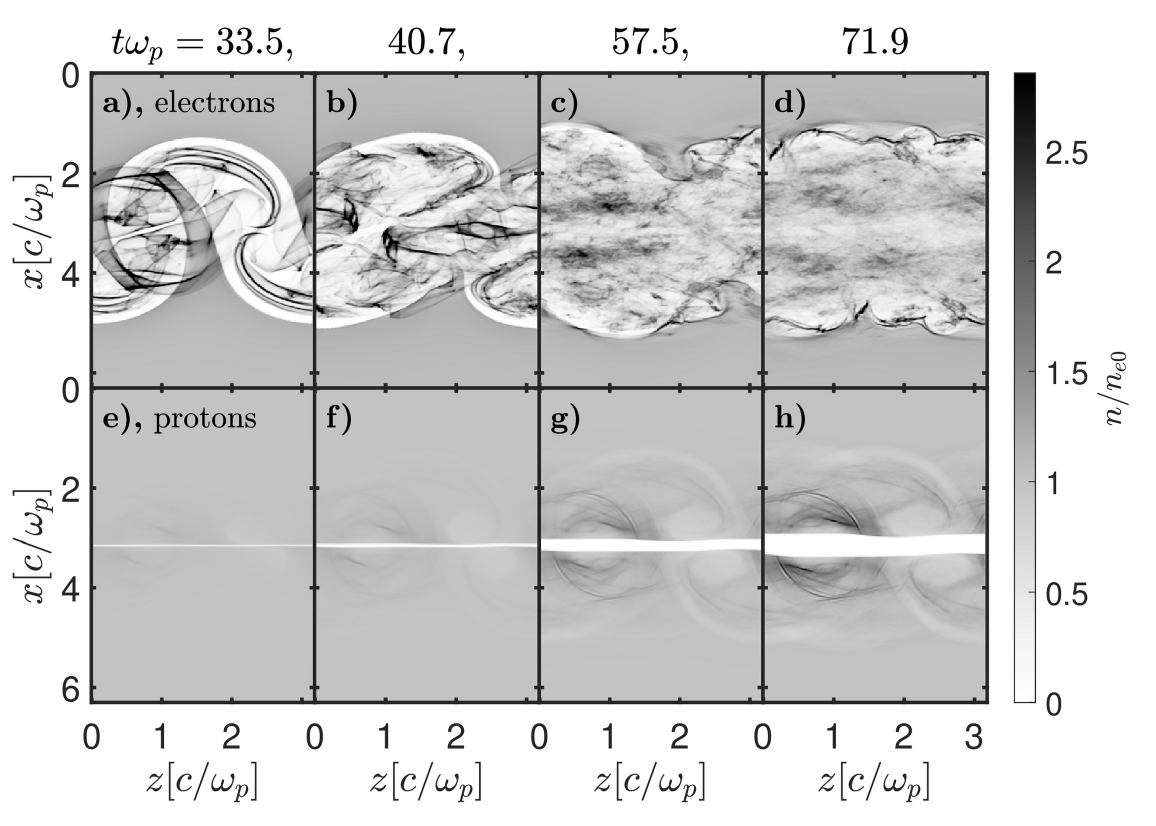}
 \caption{The evolution of the electron density [Panels (a)$\sim$(d)] and proton density [Panels (e)$\sim$(h)] in the run with $kV_0=\omega_p$ and $\omega_c=\omega_p$, at four simulation times $t\omega_p=$ 33.5 (the first column), 40.7 (the second column), 57.5 (the third column) and 71.9 (the fourth column). Only half of the simulation domain $0<x<L_x/2$ is shown.}
\label{fig:Density}
\end{figure}

In particular, the external magnetic field leads to a different morphology of the saturation stage in both runs. In the unmagnetized run, after the MI-induced magnetic field is saturated, the electron bulk continues to deform and even crosses the boundary in the $x$ direction of the simulation domain, see Fig.\ \ref{fig:Jx}(d). To reduce the effect of the finite length of the simulation domain, we have conducted another trial simulation of the unmagnetized case with 4 times the $L_x$, while the result remains similar. Still, the electron bulk deforms dramatically and breaks apart, while separate parts continue to drift in opposite directions, turning the later stage into a turbulent structure. 

Instead, in the magnetized run the ``mushroom'' as a whole gyrates under the external magnetic field $B_0 \textbf{\textit{e}}_y$, see Fig.\ \ref{fig:Jx}(h). Consequently, the deformation of the electron bulk in the $x$ direction is restricted, and a quasi-steady saturation state of the shear system is established in the magnetized run.

To better understand the saturation stage of magnetized MI, in Figure\ \ref{fig:Density} we show the evolution of electron density and proton density in the saturation stage of the $kV_0=\omega_c=\omega_p$ run. Comparing the two rows, it is obvious that the protons only begin to respond to MI and its electromagnetic field after MI enters the nonlinear stage, which is consistent with our assumption that protons remain stationary in the linear stage of MI. Note that Panels (a) and (e) are taken at the same time as in Fig.\ \ref{fig:Jx}(h). At this time, MI has entered a fully nonlinear stage [Panel (a)], while the protons only have a minimal response to the magnetic field due to their large inertia [Panel (e)], forming a narrow density gap at the shear interface. As time goes on, the ``mushrooms'' of electrons are further distorted and mix into each other, forming a turbulent structure in the gap [Panels (b) $\sim$ (d)], while the electrons outside of the gap relax to be quasi-stationary. Meanwhile, the gap in proton density continues to widen due to the strong magnetic pressure at the shear interface. Interestingly, the asymmetric structure of the magnetic field formed at an earlier time [see the white area in Panel (a)] also leaves an impact on the proton density at a later time [see the fine structure in Panel (h)]. Throughout our simulation time, the gap in proton density continues to widen. A full understanding of the ions' response requires a simulation on the ion time scale, which can be about $\sqrt{m_i/m_e}\approx42.8$ times larger than our simulation time and is beyond the scope of the current work. According to \citet{Yao2020}, in unmagnetized but thermal runs (with a much larger scale), at a later time, the expelling force of the magnetic field will be balanced by the charge separation electric field, and the gap will finally reach a maximum width. We expect a similar behavior of the ion gap from magnetized runs.

Moreover, this result suggests that MI in a planar geometry can squeeze and separate the two streams of the collisionless shear flow by forming a strong magnetic field on the shear interface. In the cylindrical geometry of jets, this effect might contribute to jet collimation or the formation of the jet spine, which has been discussed in a similar context by \citet{Kawashima2022}.

\begin{figure}[h]
\centering
\includegraphics[width=0.98 \columnwidth]{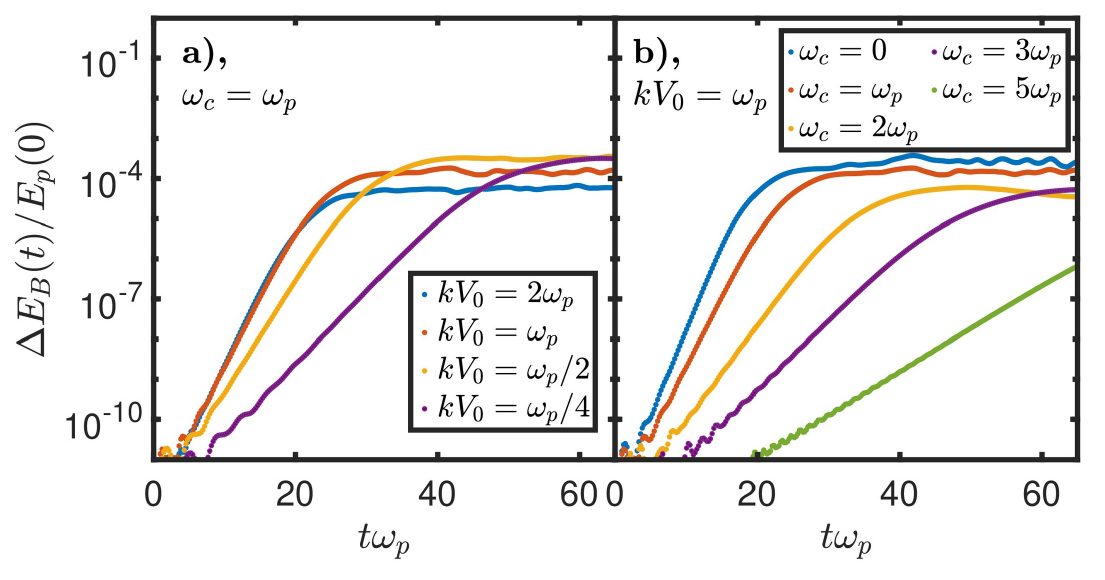}
 \caption{The evolution of the generated magnetic energy $\Delta E_B=\int dV (B-B_0)^2/2\mu_0$, divided by the initial total kinetic energy of the shear flow $E_p(0)=\int dV  n_0(m_i+m_e)c^2 (\gamma_0-1)$. Panel (a): the runs with $\omega_c=\omega_p$, and $kV_0/\omega_p=$ 2 (blue), 1 (red), 0.5 (yellow), and 0.25 (purple). Panel (b): the runs with $kV_0=\omega_p$, and $\omega_c/\omega_p=$ 0 (blue), 1 (red), 2 (yellow), 3 (purple), and 5 (green). The vertical axis is logarithmic.}
\label{fig:Benergy}
\end{figure}

We also care about the growth rate of the instability in simulations for giving a quantitative verification for our analytical results in Sec.\ \ref{sec:theo}. Figure \ref{fig:Benergy} shows the evolution of the generated magnetic energy $\Delta E_B=\int dV (B-B_0)^2/2\mu_0$ in different runs. In each run, a linear growth stage of the instability followed by a saturation stage is clearly observed. Theoretically, in the linear stage of the instability there is $\Delta E_B \propto \delta B^2\propto \exp (2\sigma t)$, so we evaluate the linear growth rate $\Gamma_s$ of $\Delta E_B$ in each run, and take $\sigma_s=\Gamma_s/2$ as the growth rate of MIs observed in simulations. The results of $\sigma_s$ are marked at corresponding wavenumbers $k_z$ in Fig.\ \ref{fig:dis05v}, where all the simulation results of $\sigma_s$ are close to the analytical dispersion relation of $\sigma(k_z)$. This is not surprising, as in a cold and collisionless plasma our equations in Sec.\ \ref{sec:setting} give a sufficiently accurate description of the plasma dynamics.

\subsection{Simulations with finite temperatures}

Although the single-mode simulations in cold plasmas reach a good agreement with our analytical results, in realistic scenarios the plasmas should be thermal less or more, which can affect the evolution of the shear interface. In consequence, we conduct simulations with finite temperatures to investigate the thermal effects on MI. Simulations are carried out with $T=1,10,100$ and 1000 eV, which corresponds to an electron thermal velocity $v_{the}$ from 0.002$c$ to 0.068$c$. We remove the initial perturbation of $v_x$ on electrons, and expect MIs to grow from thermal fluctuations spontaneously.

\begin{figure}[h]
\centering
\includegraphics[width=0.98 \columnwidth]{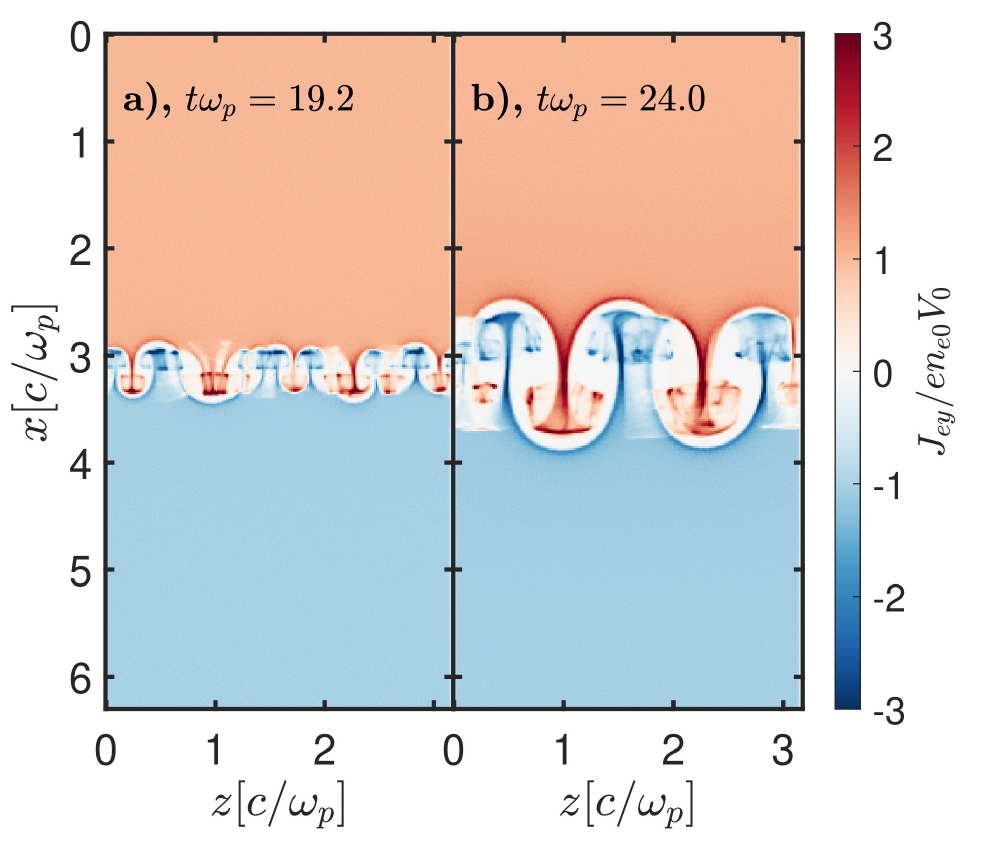}
 \caption{The evolution of the electron current $J_{ey}$, in the run with $\omega_c=0$ and $T=1$ eV, at $t\omega_p=$ 19.2 [Panel (a)] and 24.0 [Panel (b)].}
\label{fig:Jx-1ev}
\end{figure}

We confirm that in simulations, MI can grow spontaneously from fluctuations at even very low temperatures. Figure\ \ref{fig:Jx-1ev} shows the evolution of $J_{ey}$ in the $T=1$ eV unmagnetized run, where other simulation settings are the same as in the previous $kV_0=\omega_p$ single-mode run. ``Mushrooms'' with small wavelengths are excited, and merge into each other as they grow larger and enter the nonlinear growth stage. This nonlinear behavior is similar to that observed in classical Rayleigh-Taylor instabilities, although the driving mechanism here is the magnetic attraction between filaments.

\begin{figure}[h]
\centering
\includegraphics[width=0.98 \columnwidth]{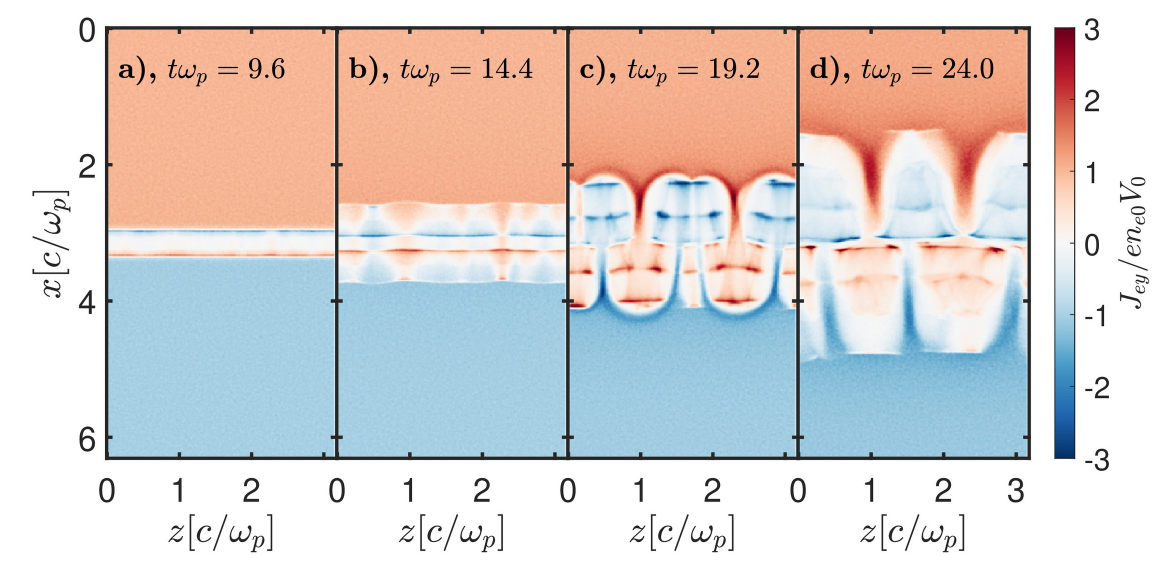}
 \caption{The evolution of the electron current $J_{ey}$, in the run with $\omega_c=0$ and $T=100$ eV, at $t\omega_p=$ 9.6 [Panel (a)], 14.4 (b), 19.2 (c) and 24.0 (d).}
\label{fig:Jx-100ev}
\end{figure}

In simulations with higher temperatures, a more different phenomenon can be resolved. Figure\ \ref{fig:Jx-100ev} shows $J_{ey}$ in the $T=100$ eV run with other settings the same. In Panel (a), on each side of the shear interface, one can observe a DC component of $J_{ey}$ in a direction opposite to the bulk current. This results from thermal diffusion of electrons across the shear interface, which cannot be described by our fluid theory but requires a kinetic approach. While ions remain stationary, the electrons crossing the interface bring a $v_y$ in the opposite direction, leading to a current imbalance on each side of the interface. Consequently, a DC component of the magnetic field $B_{\text {DC}}\sim4\sqrt{2\pi}en_0\beta_0v_{the}t$ is generated in the $z$ direction \citep{Grismayer2013DC}, which shares the same direction with the DC magnetic field formed in the saturation stage of MI [see Fig.\ \ref{fig:Jx} (d) and (h)]. Since the AC magnetic field of MI, which depends on time as $\delta \textbf{\textit{B}}\sim \textbf{\textit{B}}_{f} \exp(\sigma t)$, must grow from small thermal fluctuations $\textit{\textbf{B}}_{f}$, $B_{\text{DC}}$ will dominate at early times with the $v_{the}t$ dependence. Afterwards, the exponential dependence of MI modes will take over, forming mushroom structure from fluctuations [Panel (b)], but tainted with DC filaments [Panel (c)]. When MI enters the saturation stage, its nonlinear field appears again in the DC manner, merging into the DC magnetic field induced by thermal diffusion.

\begin{figure}[h]
\centering
\includegraphics[width=0.98 \columnwidth]{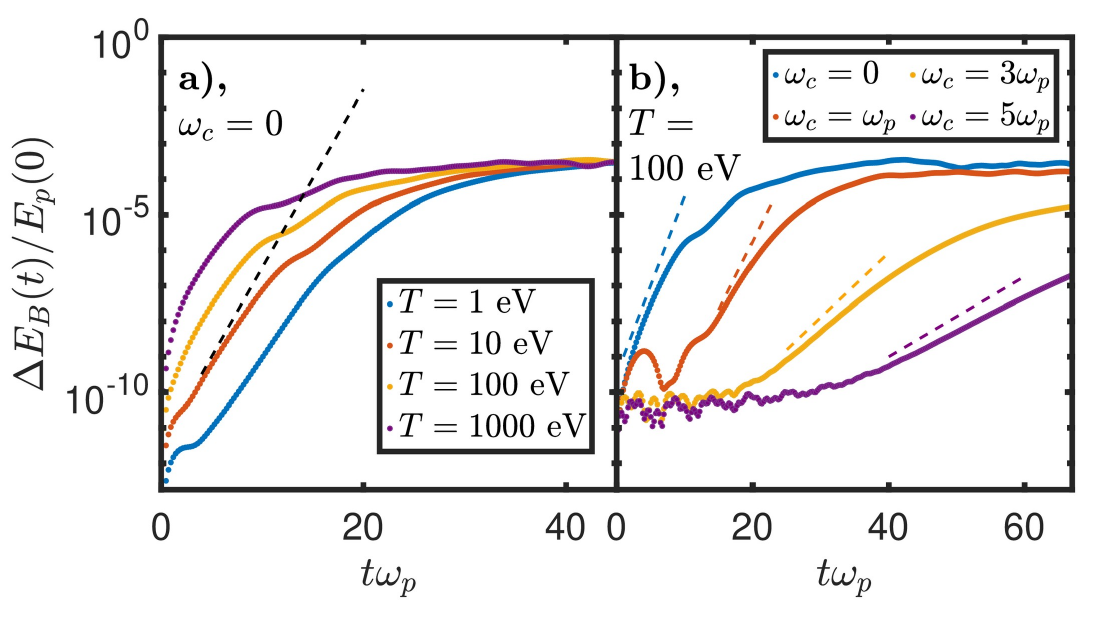}
 \caption{The same plot as Fig.\ \ref{fig:Benergy}, but in thermal simulations. Panel (a): the runs with $\omega_c=0$, and $T$ [eV]= 1 (blue), 10 (red), 100 (yellow), and 1000 (purple). $\Delta E_B/E_p\propto\exp(2\sigma_{\text{max}}t)$ with the cold plasma maximum growth rate $\sigma_{\text{max}}=\gamma_0\beta_0\omega_{p}$  is also marked as the dashed line. Panel (b): the runs with $T=100$ eV, and $\omega_c/\omega_p=$ 0 (blue), 1 (red), 3 (yellow), and 4 (purple). $\Delta E_B/E_p\propto\exp[2\sigma_{\text{max}}(\omega_c)t]$ with $\sigma_{\text{max}}$ extracted from our numerical solutions of the dispersion relation is also marked as a dash line for each $\omega_c$. 
 The vertical axis is logarithmic.}
\label{fig:Benergythermal}
\end{figure}

The competition and cooperation between the diffusion-induced DC magnetic field and the MI modes is also pronounced in the evolution of the magnetic energy. Figure\ \ref{fig:Benergythermal} shows the evolution of $\Delta E_B(t)/E_p(0)$ in the thermal runs. As seen in Panel (a), in the 1 and 10 eV runs the growth stage starts with a sharp increase that gradually smooths out, corresponding to the $v_{the}t$ dependence of the diffusion-induced DC magnetic field (note the logarithmic y-axis of the figure). Then, a linear growth stage corresponding to the MI magnetic field $B_f\exp(\sigma t)$ follows, where the order of $\Delta E_B$ increases the most. In contrast, in the 100 and 1000 eV runs, the diffusion-induced field grows faster and larger, and affects the whole growth stage more, and the linear growth stage of MI can hardly be resolved from the $\Delta E_B$ evolution.  However, at a late time, $\Delta E_B(t)/E_p(0)$ in all runs reaches about $3\times10^{-4}$ whatever the temperature, which indicates that the diffusion-induced DC field and the MI field share a similar nature. To see this directly, one can imagine superposing a perturbation in an opposite direction in Fig.\ \ref{fig:Jx}(a). This effectively gives the plasma a thermal velocity in the $x$ direction and allows the two bulk to diffuse into each other. We note that in Fig.\ \ref{fig:Jx}(a), the magnetic fields are displaced from the initial shear interface in the $x$ direction. In consequence, the superposition of an opposite mode will not cancel the fields (while in linearized fluid theory, such superposition will only lead to cancellation). Instead, their combination will appear as a DC mode of $B_z$ along the shear interface, similar to that caused by thermal diffusion of electrons. This qualitatively explains the relation between the MI fields and the thermal-induced DC fields.

Figure \ref{fig:Benergythermal}(b) shows the evolution of $\Delta E_B(t)/E_p(0)$ in the 100 eV runs under different values of $\omega_c$. The interplay between the external magnetic field and the diffusion-induced field is similar to that observed in magnetized ESKHIs previously \citep{Guo2025}, and we summarize it as follows. Under the external magnetic field, electrons diffusing across the shear interface, which are responsible for the DC field generation, will gyrate back, leading to oscillation of the DC field (see the red and yellow curves). As a result, the magnitude of the magnetic energy is restricted in the early stage, and it only begins to increase significantly when MI is excited at later time, exhibiting a linear stage much clearer than in the unmagnetized run. 

As seen in Fig.\ \ref{fig:Benergythermal}, in the linear stage of each of the thermal runs, we observe a growth rate smaller than the MI maximum growth rate predicted by our cold fluid theory. In particular, as temperature increases, the growth rate (if there is a linear stage) even slightly decreases in the unmagnetized runs [Fig.\ \ref{fig:Benergythermal}(a)]. This is contrary to the conclusion in \citet{Miller2016} that the MI growth rate increases as temperature becomes larger. This discrepancy arises because the earlier study employed a fluid approach that neglects interfacial diffusion of electrons. Although higher temperatures truly result in a larger thermal pressure and a faster dynamical response of the plasmas, in our simulations the diffusion-induced magnetic field partly magnetizes the shear interface in advance and pushes the electrons away, therefore suppressing the subsequent linear growth of MI \citep{Guo2025}.

We finally note that \citet{Alves2015Transverse} concluded that in the relativistically thermal regime ($T\gtrsim50$ keV), the increase in temperature will also lead to a decreased growth rate. Despite the similarity between our conclusions, they are very different in essence; The decrease in our context happens due to the difference in thermal velocities of electrons and protons, while in \citet{Alves2015Transverse} it happens due to an enhanced relativistic particle mass. The latter work has confirmed that MI with a decreased growth rate happens in simulations of relativistically thermal electron-positron shear flows. Instead, in relativistically thermal electron-proton shear flows, we will expect more of a strong diffusion-induced DC field, which completely suppresses the later growth of MI. Due to a much larger requirement on computation resource, we did not investigate this scenario in this paper and leave it for future works.

\section{Conclusion} \label{sec:conclusion}
In this paper, we have investigated the effect of a flow-aligned magnetic field on MI, with both theoretical analyses and PIC simulations. In the theoretical part, we consider a generic shear flow under a uniform magnetic field, and analyze the system with two-fluid equations in the limit of a cold and collisionless plasma. In this way, we derive the eigenequations for the linear growth rates of MI. With a extension on the perturbation wavevector, our eigenequation can be generalized to describe not only MI (with $\textbf{\textit{k}}$ perpendicular to the shear), but also ESKHI ($\textbf{\textit{k}}$ parallel to the flow) and a more general coupled instability with an arbitrary wavevector $\textbf{\textit{k}}$ in the plane of the shear flow.

Numerical solutions of the MI dispersion relation suggest that the external flow-aligned magnetic field always has a suppressing effect on MI. However, compared with ESKHI, MI is more robust against the external magnetic field, and within our framework there appears to be no stabilizing threshold magnetic field for MI; In other words, MI remains unstable under very strong magnetic fields. In consequence, we expect MI to dominate the dynamics of the shear flow not only in relativistic scenarios, but also in strongly magnetized sub-relativistic scenarios. We also investigate a few cases of the coupled instability with $\textit{\textbf{k}}$ over the whole plane. The result reveals that in shear flows with $V_0\sim c/3$, the coupling can lead to a dynamics different from a pure ESKHI or MI, and the external magnetic field can be destabilizing for some wavevectors. In contrast, in relativistic scenarios the instability growth rate still peaks at $k_y=0$, which corresponds to a pure MI.

In the simulation part, we investigate MI in a planar and collisionless shear flow with PIC simulations. First, we conduct single-mode simulations in cold plasmas ($v_{th}=0$) under different magnitudes of external magnetic fields. Results of the simulations reach a good agreement with our theoretical results, where growth rates extracted from the simulations are very close to our numerical solution of the MI dispersion relation. In unmagnetized runs, we observe the formation of the mushrooms and a dramatical deformation of the flow bulk in the saturation stage. In contrast, in magnetized runs the mushrooms are redirected to gyrate in the plane, restricting the deformation of the flow bulk. Especially, a quasi-steady density gap is formed in the magnetized runs, where electrons and protons are pushed away from the shear interface, indicating the potential of magnetized MI in jet spine collimation.

Simulations with finite temperatures are also carried out. In thermal simulations, MI growth spontaneously from thermal fluctuations, and the ``mushrooms'' merge into each other as they enter the nonlinear stage. As temperature increases, a diffusion-induced DC magnetic field occurs on the shear interface, which shares a similar morphology with the MI saturation field. This DC field grows faster than MI and dominates the early stage of the shear flow. In shear flows with higher temperatures, the DC field can reach a high magnitude, thereby skipping or suppressing the linear growth phase of MI.

We note that all our analytical results are derived in the limit of a cold and collisionless plasma, so their applicability on thermal jets remains to be investigated. Our simulations have shown that even a low temperature can lead to very different dynamics of the shear flow, especially on the shear interface. Generally speaking, the combination of different temperature and different particle species (protons and positrons) can leave different implications in realistic scenarios, which is worth exploring in the future.

Observations have revealed that the magnetic field structure can be actually helical in astrophysical jets \citep{Hovatta_2012}. How this magnetic field interacts, or originates from some mechanism associated with MI, remains a critical open question. Moreover, how the cylindrical geometry of jets affects MI remains theoretically unclear. Since MI has only been discovered recently \citep{Alves2015Transverse} and received limited attention in mainstream jet research, further work is needed to bridge this plasma instability from simplified theory and simulations to large-scale and realistic jet dynamics.

\begin{acknowledgments}
This work was supported by the Strategic Priority Research Program of Chinese Academy of Sciences (Grant No. XDA25010100 and XDA25050500), Science Challenge Project (No. TZ2025012), National Natural Science Foundation of China (Grant No. 12075204) and Shanghai Municipal Science and Technology Key Project (Grant No. 22JC1401500). Dong Wu thanks the sponsorship from Yangyang Development Fund.
\end{acknowledgments}

%






\appendix

\section{generalized shear instability analyses}
\label{app:3D}

In this appendix we show the linearization process of Eqs.\ \ref{eq:1} $\sim$ \ref{eq:5} in detail. The physical settings of the shear flow are the same as in Sec.\ \ref{sec:setting}, but the perturbations are generalized to be in the form of $f(x,y,z,t)=\tilde{f}(x)e^{i(k_yy+k_zz-\omega t)}$. The linearized form of Eqs.\ \ref{eq:1} and \ref{eq:2} is:
\begin{align}
&\tilde n_e=\frac{-i}{\Omega}\left[\frac{\partial }{\partial x}(\tilde v_xn_{e0})+in_{e0}(k_y\tilde v_y+k_z\tilde v_z)\right],\label{eq:n1}\\
&    \tilde v_x=\frac{e}{\gamma_0 m_e}\frac{-i}{\Omega}(\tilde E_x+v_0\tilde B_z-\tilde v_zB_0), \label{eq:vx} \\    
&    \tilde v_y=\frac{e}{\gamma_0^3 m_e}\frac{-i}{\Omega}\left[\tilde E_y+\frac{m_e}{e}\tilde v_x\frac{\partial}{\partial x}(\gamma_0 v_0)\right],\label{eq:vy} \\
&    \tilde v_z=\frac{e}{\gamma_0 m_e}\frac{-i}{\Omega}(\tilde E_z-v_0\tilde B_x+\tilde v_xB_0),  \label{eq:vz}  
\end{align}
where $\Omega\equiv\omega-k_yv_0$. We can decouple $\tilde v_x$ and $\tilde v_z$ from Eqs.\ \ref{eq:vx} and \ref{eq:vz}:
\begin{align}
&    \tilde v_x=-\frac{e}{\gamma_0 m_e}\frac{\omega_c (-\tilde E_z+\tilde B_xv_0)+i\Omega(\tilde E_x+\tilde B_zv_0)}{\Omega^2-\omega_c^2},\label{eq:vx1} \\
& \tilde v_z=-\frac{e}{\gamma_0 m_e}\frac{\omega_c(\tilde E_x+\tilde B_zv_0)+i\Omega(\tilde E_z-\tilde B_xv_0)}{\Omega^2-\omega_c^2}.\label{eq:vz1}
    \end{align}
Now we linearize Eq.\ \ref{eq:5}:
\begin{align}
  &  \tilde J_x=\frac{e^2n_{e0}}{\gamma_0m_e}\frac{\omega_c (-\tilde E_z+\tilde B_xv_0)+i\Omega(\tilde E_x+\tilde B_zv_0)}{\Omega^2-\omega_c^2 },\label{eq:Jx}\\
  &  \tilde J_y=\frac{e^2n_{e0}}{\gamma_0^3 m_e}\frac{i\omega}{\Omega^2}\tilde E_y+ie\frac{\partial}{\partial x}\left(\frac{v_0\tilde v_xn_{e0}}{\Omega}\right)-\frac{ev_0n_{e0}}{\Omega}k_z\tilde v_z,\label{eq:Jy} \\
  &  \tilde J_z=\frac{e^2n_{e0}}{\gamma_0m_e}\frac{\omega_c (\tilde E_x+\tilde B_zv_0)+i\Omega(\tilde E_z-\tilde B_xv_0)}{\Omega^2-\omega_c^2 },\label{eq:Jz}
\end{align}
and Eq.\ \ref{eq:3}:
\begin{align}
&\tilde B_x=\frac{1}{\omega}(k_y\tilde E_z-k_z\tilde E_y), \label{eq:Bx}\\
&\tilde B_z=-\frac{i}{\omega}\frac{\partial \tilde E_y}{\partial x}-\frac{k_y}{\omega}\tilde E_x,\label{eq:Bz}
\end{align}
Replace $\textbf{\textit{B}}$ in Eq.\ \ref{eq:4} with Eqs.\ \ref{eq:Bx} and  \ref{eq:Bz}, we can derive
\begin{align}
ik_y\frac{\partial \tilde E_y}{\partial x}+ik_z\frac{\partial \tilde E_z}{\partial x}=i\omega \mu_0 \tilde J_x+\left(\frac{\omega^2}{c^2}-k_y^2-k_z^2\right)\tilde E_x,\label{eq:MEx}\\
ik_y\frac{\partial \tilde  E_x}{\partial x}-\frac{\partial^2 \tilde E_y}{\partial x^2}-k_yk_z\tilde E_z=i\omega \mu_0 \tilde J_y+\left(\frac{\omega^2}{c^2}-k_z^2\right)\tilde E_y,\label{eq:MEy} \\
ik_z\frac{\partial \tilde E_x}{\partial x}-\frac{\partial^2 \tilde E_z}{\partial x^2}-k_yk_z\tilde E_y=i\omega \mu_0 \tilde J_z+\left(\frac{\omega^2}{c^2}-k_y^2\right)\tilde E_z. \label{eq:MEz}   
\end{align}
Since $\tilde J_x$ only contains first order terms of $\tilde E_x$, from Eq.\ \ref{eq:MEx} we can solve $\tilde E_x$ in terms of only two perturbed variables $\tilde E_y$ and $\tilde E_z$:
\begin{eqnarray}
\tilde E_x=&&\frac{i\left[k_yc^2\left(\Omega^2-\omega_c^2 \right)-\gamma_0^2\omega_p^2\Omega v_0\right]}{D(\omega,k)} \frac{\partial \tilde E_y}{\partial x}+\frac{i k_zc^2\left(\Omega^2-\omega_c^2 \right) }{D(\omega,k)} \frac{\partial \tilde E_z}{\partial x}+\frac{i\gamma_0^2\omega_p^2\omega_ck_zv_0}{D(\omega,k)} \tilde E_y
+\frac{i\gamma_0^2\omega_p^2\omega_c\Omega}{D(\omega,k)} \tilde E_z, \label{eq:Ex}
\end{eqnarray}
where $D(\omega,k)=\Omega^2\left[\omega^2-(k_y^2+k_z^2)c^2-\gamma_0^2\omega_p^2\right]-\omega_c^2[\omega^2-(k_y^2+k_z^2)c^2]$ is the generalization of $D(\omega,k_z)$ defined in Sec.\ \ref{sec:eigen}. Now all perturbed variables in the system can be expressed as functions of $\tilde E_y,\tilde E_z$  and their partial derivatives with respect to $x$.
Substitute them into Eqs.\ \ref{eq:MEy} and \ref{eq:MEz}, we finally derive the two coupled eigenequations of the system:
\begin{eqnarray}
       &&\left[\frac{\omega^2(\omega_c^2+\omega_p^2-\Omega^2)+k_z^2c^2(\Omega^2-\omega_c^2+(\gamma_0^2-1)\omega_p^2)}{D(\omega,k)}\tilde E_y'\right]'
       -\left[\frac{k_z\omega_c\omega_p^2[(1-\gamma_0^2)\omega^2+\gamma_0^2(k_z^2v_0^2+k_yv_0\omega)]}{D(\omega,k)\Omega}\right]'\tilde E_y\nonumber\\
       &&+\left[\frac{\omega^2}{c^2}\left(\frac{\omega_p^2}{\Omega^2}-1 \right)+\frac{k_z^2v_0^2\gamma_0^2\omega_p^2}{c^2(\Omega^2-\omega_c^2)}+k_z^2\right]\tilde E_y
      +\left[\frac{k_z(v_0\Omega \gamma_0^2\omega_p^2-k_yc^2(\Omega^2-\omega_c^2))}{D(\omega,k)}\tilde E_z'\right]' \nonumber\\
       &&+\left[\frac{\omega_c\gamma_0^2\omega_p^2\left(\omega^2v_0/c^2-\omega k_y-k_z^2v_0\right)}{D(\omega,k)}\tilde E_z\right]'+
       \frac{k_z^2v_0\gamma_0^2\omega_c\omega_p^2}{D(\omega,k)}\tilde E_z'+
       \left[\frac{\Omega k_zv_0\gamma_0^2\omega_p^2}{c^2(\Omega^2-\omega_c^2)}-k_y k_z\right] \tilde E_z=0 ,\label{eq:disa1}  
       \end{eqnarray}
       
\begin{eqnarray}
   && \left[\left(1+\frac{c^2k_z^2(\Omega^2-\omega_c^2)}{D(\omega,k)}\right)\tilde E_z'\right]'
    +\left[\frac{k_z\gamma_0^2\Omega\omega_c\omega_p^2}{D(\omega,k)}\right]'\tilde E_z 
    +\frac{1}{c^2}\frac{F(\omega,k)-c^2k_z^2[(c^2k^2-\omega^2)(\Omega^2-\omega_c^2)+\gamma_0^2\Omega^2\omega_p^2]}{D(\omega,k)}\tilde E_z \nonumber\\
    &&+\left[\frac{k_z(c^2k_y(\Omega^2-\omega_c^2)-v_0\gamma_0^2\Omega\omega_p^2)}{D(\omega,k)}\tilde E_y'\right]'
    +\left[\frac{k_z^2v_0\gamma_0^2\omega_c\omega_p^2}{D(\omega,k)}\tilde E_y\right]' \nonumber\\
    &&-\frac{\omega_c\gamma_0^2\omega_p^2\left(k_y\omega+k_z^2v_0-\omega^2 v_0/c^2\right)}{D(\omega,k)}\tilde E_y'
    +\left[\frac{k_zv_0\Omega\gamma_0^2\omega_p^2(c^2k^2-\omega^2+\gamma_0^2\omega_p^2)}{c^2D(\omega,k)}+k_yk_z\right]\tilde E_y=0,  \label{eq:disa2}
\end{eqnarray}
where $F(\omega,k)\equiv\Omega^2[\omega^2-(k_y^2+k_z^2)c^2-\gamma_0^2\omega_p^2]^2-\omega_c^2[\omega^2-(k_y^2+k_z^2)c^2]^2$. Still, the connecting boundary conditions are  derived by integrating Eqs.\ \ref{eq:disa1} and \ref{eq:disa2} across the shear interface (e.g. $x=0$):
\begin{eqnarray}
       &&\left[\frac{\omega^2(\omega_c^2+\omega_p^2-\Omega^2)+k_z^2c^2(\Omega^2-\omega_c^2+(\gamma_0^2-1)\omega_p^2)}{D(\omega,k)}\tilde E_y'\right]\bigg |_{0_-}^{0_+}
       -\left[\frac{k_z\omega_c\omega_p^2[(1-\gamma_0^2)\omega^2+\gamma_0^2(k_z^2v_0^2+k_yv_0\omega)]}{D(\omega,k)\Omega}\tilde E_y\right]\bigg |_{0_-}^{0_+}\nonumber\\
       &&      +\left[\frac{k_z(v_0\Omega \gamma_0^2\omega_p^2-k_yc^2(\Omega^2-\omega_c^2))}{D(\omega,k)}\tilde E_z'\right]\bigg |_{0_-}^{0_+} +\left[\frac{\omega_c\gamma_0^2\omega_p^2\left(\omega^2v_0/c^2-\omega k_y-k_z^2v_0\right)}{D(\omega,k)}\tilde E_z\right]\bigg |_{0_-}^{0_+}=0,\label{eq:conna1}  
       \end{eqnarray}
       
\begin{eqnarray}
   && \left[\left(1+\frac{c^2k_z^2(\Omega^2-\omega_c^2)}{D(\omega,k)}\right)\tilde E_z'\right]\bigg |_{0_-}^{0_+}
    +\left[\frac{k_z\gamma_0^2\Omega\omega_c\omega_p^2}{D(\omega,k)} \tilde E_z\right]\bigg |_{0_-}^{0_+} 
  \nonumber\\
    &&+\left[\frac{k_z[c^2k_y(\Omega^2-\omega_c^2)-v_0\gamma_0^2\Omega\omega_p^2]}{D(\omega,k)}\tilde E_y'\right]\bigg |_{0_-}^{0_+}
    +\left[\frac{k_z^2v_0\gamma_0^2\omega_c\omega_p^2}{D(\omega,k)}\tilde E_y\right]\bigg |_{0_-}^{0_+}=0,  \label{eq:conna2}
\end{eqnarray}
If we set $k_y=0$, the equations are reduced to Eqs.\ \ref{eq:dis1}$\sim$\ref{eq:dis20} derived in Sec.\ \ref{sec:eigen}. In contrast, if we set $k_z=0$, the equations are reduced to the dispersion relation of magnetized ESKHI derived in \citet{Guo2025}.

When $\omega_c\neq0$ and the zeroth-order variables are uniform on each side of the shear interface, the equations can be solved by assuming solutions with the form of $\tilde E_y^{\pm}=A_1^{\pm}\exp(a^{\pm}x)+A_2^{\pm}\exp(b^{\pm}x)$ and $\tilde E_z^{\pm}=B_1^{\pm}\exp(a^{\pm}x)+B_2^{\pm}\exp(b^{\pm}x)$. A detailed introduction of this scheme can be found in the Appendix in \citet{Guo2025}.

Instead, in the unmagnetized case with $\omega_c=0$, Eqs.\ \ref{eq:disa1} and \ref{eq:disa2} become 
\begin{eqnarray}
       &&\left[\frac{\omega^2(\omega_p^2-\Omega^2)+k_z^2c^2(\Omega^2+(\gamma_0^2-1)\omega_p^2)}{\Omega^2\left(\omega^2-(k_y^2+k_z^2)c^2-\gamma_0^2\omega_p^2\right)}\tilde E_y'\right]'
       +\left[\frac{\omega^2(\omega_p^2-\Omega^2)+k_z^2c^2(\Omega^2+(\gamma_0^2-1)\omega_p^2)}{c^2\Omega^2}\right]\tilde E_y   \nonumber\\
     && +\left[\frac{k_z(v_0\Omega \gamma_0^2\omega_p^2-k_yc^2\Omega^2)}{\Omega^2\left[\omega^2-(k_y^2+k_z^2)c^2-\gamma_0^2\omega_p^2\right]}\tilde E_z'\right]' +
       \left[\frac{k_z(v_0\Omega \gamma_0^2\omega_p^2-k_yc^2\Omega^2)}{c^2\Omega^2}\right] \tilde E_z=0 ,
       \end{eqnarray}
\begin{eqnarray}
   && \left[\frac{\omega^2-k_y^2c^2-\gamma_0^2\omega_p^2}{\omega^2-(k_y^2+k_z^2)c^2-\gamma_0^2\omega_p^2}\tilde E_z'\right]'
    +\frac{\omega^2-k_y^2c^2-\gamma_0^2\omega_p^2}{c^2}\tilde E_z \nonumber\\
    &&+\left[\frac{k_z(c^2k_y\Omega^2-v_0\gamma_0^2\Omega\omega_p^2)}{\Omega^2(\omega^2-(k_y^2+k_z^2)c^2-\gamma_0^2\omega_p^2)}\tilde E_y'\right]'
    +\left[\frac{k_z(c^2k_y\Omega^2-v_0\gamma_0^2\Omega\omega_p^2)}{c^2\Omega^2}\right]\tilde E_y=0,  
\end{eqnarray}
where one can easily find that the solution to $\tilde E_y$ and $\tilde E_z$ is the single mode $C_i\exp\left[- |x|\sqrt{(k_y^2+k_z^2)+(\gamma_0^2\omega_p^2-\omega^2)/c^2}\right]$. Substitute the solution into Eqs.\ \ref{eq:conna1} and \ref{eq:conna2}, and one can derive the dispersion relation of the unmagnetized shear flow. The special case of unmagnetized ESKHI (with $k_z=0$) and MI (with $k_y=0$) can be solved analytically.

A simple numerical implementation of the above scheme can be found at \url{https://github.com/YaoGuo2001/MI/}. The notebook therein can be used to solve the eigenfrequency of a symmetrical shear flow with a uniform density, with arbitrary values of $V_0,k_y,k_z$ and $B_0$.

\bibliography{sample631}{}
\bibliographystyle{aasjournal}



\end{document}